\documentclass{article}

\usepackage{arxiv}

\usepackage[utf8]{inputenc} 
\usepackage[T1]{fontenc}    
\usepackage{hyperref}       
\usepackage{url}            
\usepackage{booktabs}       
\usepackage{amsfonts}       
\usepackage{nicefrac}       
\usepackage{microtype}      
\usepackage{lipsum}		
\usepackage{graphicx}
\usepackage{doi}
\usepackage{threeparttablex}
\usepackage{sistyle}
\SIthousandsep{,}

\usepackage{import}

\def\bZ{{\mathbf Z}}
\def\bbeta{{\boldsymbol{\beta}}}

\usepackage{amsthm}

\usepackage{booktabs}
\usepackage{amsmath}

\newcommand{\E}{\textrm{E}}
\newcommand{\betahat}{\hat{\beta}}
\newcommand{\bbetahat}{\hat{\pmb{\beta}}}
\newcommand{\alphahat}{\hat{\alpha}}
\newcommand{\balphahat}{\hat{\pmb{\alpha}}}
\newcommand{\sigmahat}{\hat{\sigma}}
\newcommand{\Xtilde}{\widetilde{X}}
\newcommand{\Vhat}{\widehat{\textrm{V}}}
\newcommand{\simtabnote}{\textbf{Bias} and \textbf{ESE} are, respectively, the empirical relative bias and standard error of the log prevalence ratio estimator $\betahat_1$; \textbf{ASE} is the average of the standard error estimator $\widehat{\textrm{SE}}(\betahat_1)$; \textbf{CP} is the empirical coverage probability of the 95\% confidence interval for the log prevalence ratio $\beta_1$; \textbf{RE} is the empirical relative efficiency to the Gold Standard. All entries are based on \num{1000} replicates.}
\usepackage{mathtools}
\DeclarePairedDelimiter{\nint}\lfloor\rceil 
\usepackage{dsfont}

\title{Combining straight-line and map-based distances to investigate the connection between proximity to healthy foods and disease}

\author{ \href{https://orcid.org/0000-0001-5380-2427}{\includegraphics[scale=0.06]{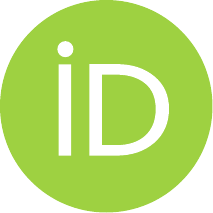}\hspace{1mm}Sarah C.~Lotspeich} \\
	Department of Statistical Sciences\\
	Wake Forest University\\
	Winston-Salem, NC 27109 \\
	\texttt{lotspes@wfu.edu} \\
	\And
	Ashley E.~Mullan \\
	Department of Statistical Sciences\\
	Wake Forest University\\
	Winston-Salem, NC 27109 \\
	\texttt{mullae22@wfu.edu} \\
        \And
        Lucy~D'Agostino McGowan\\
        Department of Statistical Sciences\\
	Wake Forest University\\
	Winston-Salem, NC 27109 \\
	\texttt{mcgowald@wfu.edu} \\
        \And
        Staci A.~Hepler \\
        Department of Statistical Sciences\\
	Wake Forest University\\
	Winston-Salem, NC 27109 \\
	\texttt{heplersa@wfu.edu} \\
}

\renewcommand{\shorttitle}{Combining straight-line and map-based distances}

\hypersetup{
pdftitle={Food access imputation},
pdfsubject={q-bio.NC, q-bio.QM},
pdfauthor={Sarah C.~Lotspeich, Ashley E.~Mullan, Lucy~D'Agostino McGowan, Staci A.~Hepler},
pdfkeywords={census tract, diabetes, Google Maps, Haversine formula, measurement error, missing data, network distance, obesity, SNAP},
}

\begin{document}
\maketitle

\begin{abstract}
Healthy foods are essential for a healthy life, but accessing healthy food can be more challenging for some people than others. This disparity in food access may lead to disparities in well-being, potentially with disproportionate rates of diseases in communities that face more challenges in accessing healthy food (i.e., low-access communities). Identifying low-access, high-risk communities for targeted interventions is a public health priority, but current methods to quantify food access rely on distance measures that are either computationally simple (like the length of the shortest straight-line route) or accurate (like the length of the shortest map-based driving route), but not both. We propose a multiple imputation approach to combine these distance measures, allowing researchers to harness the computational ease of one with the accuracy of the other. The approach incorporates straight-line distances for all neighborhoods and map-based distances for just a subset, offering comparable estimates to the ``gold standard'' model using map-based distances for all neighborhoods and improved efficiency over the ``complete case'' model using map-based distances for just the subset. Through the adoption of a measurement error framework, information from the straight-line distances can be leveraged to compute informative placeholders (i.e., impute) for any neighborhoods without map-based distances. Using simulations and data for the Piedmont Triad region of North Carolina, we quantify and compare the associations between two health outcomes (diabetes and obesity) and neighborhood-level access to healthy foods. The imputation procedure also makes it possible to predict the full landscape of food access in an area without requiring map-based measurements for all neighborhoods.
\end{abstract}


\section{Motivation}
\label{sec:motivation}

Eating healthy foods is critical to childhood development and preventing illnesses in adulthood, including diabetes and obesity. \cite{Liu2000, Harding2008} However, eating healthy foods is not just a choice; having access to healthy food options also determines what people eat. For some people, predominantly those with disabilities and in low-income or minority communities, healthy foods are not always accessible. \cite{Bower2014, Brucker2017} Access can be hampered by physical factors, like geographic proximity and the availability of public transit, \cite{Bennion2022} and social factors, like structural racism and discrimination. \cite{Burke2018, Odoms-Young&Bruce2018} Thus, not just consumption of healthy foods but access to them can impact health, \cite{Ahern2011, Gallegos2021, Kanchi2021, Li2021} and disparities in access can potentially perpetuate disparities in health. \cite{Larson2009} 

The United States Department of Health and Human Services defines five domains of social determinants of health (SDOH): (i) economic stability, (ii) education access and quality, (iii) health care access and quality, (iv) neighborhood and built environment, and (v) social and community context. \cite{HealthyPeople2030} These domains cover various aspects of peoples' surroundings (home, work, etc.) that can affect health outcomes. Access to healthy foods falls primarily under domain (iv) since it hinges upon the presence of healthy food retailers, like grocery stores and farmer's markets, in a person's environment. Food insecurity is an important, related SDOH, defined as ``a household-level economic and social condition of limited or uncertain access to adequate food.'' \cite{USDA_foodinsecurity} Importantly, a food-insecure household can still have access since the former depends on economic stability (domain i) while the latter is a function of their built environment (domain iv). Here, we focus on food access as an SDOH.

For public health officials seeking to reduce the burden of diet-related non-communicable diseases, understanding (i) the landscape of food access in a community and (ii) the disease's connection to food access can be informative. When helping public health officials understand these issues, quantifying food access is an important place to start. To identify low-access, high-risk neighborhoods, food access can be quantified using (i) \textit{proximity}, i.e., how far away healthy food stores are, or (ii) \textit{density}, i.e., how many healthy food stores are within some radius. Many other food access metrics are also possible, including a binary indicator of at least one healthy food store within some radius, the relative density of healthy to unhealthy food stores within some radius, or the average distance to healthy food stores within some radius. Using geocoded store locations, we focus on neighborhood proximity to healthy foods, defined as the minimum distance between a neighborhood's ``population center'' (a population-weighted centroid) and a healthy food store, and the connection between proximity to healthy foods and diagnosed diabetes and obesity prevalence. We chose proximity to healthy foods for its granular ability to quantify access (in miles) rather than as a discretized count or indicator. This choice also avoided the need to specify a buffer zone, and analyses can be sensitive to how these zones are defined (e.g., a $1$-mile buffer versus a $0.5$-mile buffer can lead to different conclusions). \cite{James2014}

\subsection{Drawbacks of current distance calculations}

When calculating distances between neighborhoods and healthy food stores, there currently exists a trade-off between (i) computationally simple methods that are less accurate and (ii) more accurate methods that are computationally complex. Euclidean distance, which draws a straight line along the shortest path between two points (the healthy food store location and the neighborhood's population center) ``as the crow flies'' is among the former. A recent systematic review of studies in the United Kingdom found that Euclidean distance and circular buffers based on it were the most commonly used methods to quantify food access. \cite{Titis2021} (Buffers are commonly circular catchment areas of some specified radius, e.g., one Euclidean mile, from a point of interest.) They are also prominent in studies of U.S. food access. \cite{Peng2020, Sanchez2022} The Haversine formula, which instead draws the shortest possible arc between two points, can offer better accuracy, particularly over long distances, and it is also computationally simple. Still, Euclidean and Haversine calculations ignore transportation infrastructure (e.g., roads or public bus/train routes) and natural barriers (e.g., rivers or mountains), underestimating proximity to healthy foods and overestimating neighborhood access. These problems could be exacerbated in rural areas with fewer available stores and fewer accessible straight-line routes.

Map-based (or network) distances more accurately quantify neighborhoods’ proximity to healthy foods and more realistically describe the landscape of food access, \cite{Bennion2022} and they can differ substantially from straight-line ones. For the example in Supplemental Figure~S1, it can be seen that the estimated driving distance between the historic Reynolda House in Winston-Salem, North Carolina (NC) and a nearby Food Lion grocery store ($2.65$ miles) was $1.6$ times the straight-line distance between them ($1.64$ miles). Further, the straight-line route in Supplemental Figure~S1 cuts through a pond, making it impassible. 

Improved versions of the food access measures like proximity can be calculated from map-based distances. \cite{Sharkey2009, Matsuzaki2023} The Google Maps API is incredibly powerful for calculations like this, and open-source software in the statistical computing language R can integrate the API into a statistical workflow. \cite{R, KahleWickham2013, Cooley2023} These tools can calculate the map-based travel distances, offering a more accurate snapshot of a neighborhood’s access. While we focus on map-based driving distances, map-based distances based on walking or public transportation can also be obtained. These driving distances were calculated from the fastest route under ``normal'' traffic conditions, although the software also offers further customization (e.g., a 5:00 PM ``rush hour'' departure time). Still, making all necessary calculations may not be feasible due to time-intensive computations and monthly limits on free API usage. Even paid options like ArcGIS can quickly become time-intensive, so obtaining map-based distances for entire studies is potentially unrealistic at scale (e.g., to shape policy at the state or national level versus the county). 

\subsection{Straight-line food access as a mismeasured covariate}\label{intro:errors}

With fixed resources, researchers may have to choose between a small-scale, more accurate study or a large-scale, less accurate one. For example, it may be reasonable to obtain map-based distances for all neighborhoods in one city or county but not in the state. Instead, researchers might sacrifice accuracy to broaden the scope of their study, leading to biased estimates and statistical power loss in the downstream analysis. \cite{Carroll2006} Still, what if accurate, map-based distances can be calculated for some neighborhoods? 

The two-phase design, or internal validation study, is commonly used in the measurement error literature and can be applied here. \cite{White1982, Giganti2020, Nab2021, Lotspeich2022} Straight-line distances are collected for all neighborhoods in Phase I. Then, in Phase II, map-based distances are collected for a chosen subset, which can be strategically selected based on any information fully observed in Phase I. For example, Phase II in the Piedmont Triad analysis was chosen via county-stratified random sampling to ensure geographic diversity. Many other sampling designs are possible, including strategic ones to target different goals (e.g., to minimize the variance of the downstream model estimates). 

Straight-line and map-based methods seek to measure the same quantity but take different paths to get there. Thus, assuming that straight-line distance underestimates map-based, a measurement error model is a natural way to relate more and less accurate food access variables derived from these distances (like proximity). Let $X$ and $X^*$ denote a neighborhood's proximity to healthy foods based on map-based (error-free) and straight-line (error-prone) distances, respectively. The classical measurement error model is a popular choice: $X^* = X + U$, where the random errors $U$ are assumed to be independent of $X$. \cite{Fuller1987} Often, $U$ is assumed to follow a mean-zero normal distribution, but when measuring proximity to healthy foods, the straight-line measure $X^*$ systematically underestimates map-based $X$ by some absolute amount, so $U \leq 0$ necessarily and a different distribution is needed. It may also be desirable to model straight-line $X^*$ as systematically underestimating map-based $X$ by some relative factor instead. A multiplicative error model is also possible: $X^* = XW$,  where $W$ is once again assumed to be independent of $X$. Multiplicative measurement error models are not as common as additive ones, and there is no most popular distribution to assume for them. The main consideration in selecting a distribution for $W$ is that $0 \leq W \leq 1$, since $X^* \leq X$. Fortunately, we do not need to specify the measurement error model to use the proposed multiple imputation approach, and the methods work well under either model. In treating the straight-line proximity to healthy foods as an error-prone version of the map-based proximity, a manageable statistical problem replaces an unmanageable computational one. 

\subsection{Statistical modeling with a mismeasured covariate}

First, suppose that the errors in the straight-line distances were ignored. If $X^*$ were used instead of $X$ to conduct the so-called ``naive'' analysis of food access and disease prevalence, the model estimates would be biased and inference underpowered. \cite{Carroll2006} Linear regression estimates will be attenuated (i.e., biased toward zero), \cite{Fuller1987} while logistic and Poisson regression estimates can be attenuated or exaggerated. \cite{Barron1977, VidekWong1996} These results echo intuition: Measurement errors must be addressed and analyses corrected to obtain valid inferences about the relationship between food access and health. 

In the two-phase study, $X^*$ is observed for all neighborhoods and $X$ for only the chosen subset. Many statistical methods can be used to analyze the resulting data with no missing $X^*$ but some missing $X$, and the popular ones fall into two groups: (i) design-based and (ii) model-based estimators. Most methods require at least partial information on the relationship between error-prone and error-free variables, and these two groups are named for how they handle the missing data in $X$ left behind from the validation study. Complete data from the validation study can be leveraged to overcome the missing map-based measures.

Design-based estimators rely primarily on the sampling probabilities describing which neighborhoods were chosen for map-based measurements (i.e., the study design) to overcome the missing data in $X$. This class includes the popular inverse probability weighted (IPW) \cite{HorvitzThompson1952} and augmented IPW/generalized raking estimators. \cite{Deville1993, RobinsRotnitzky&Zhao1994, OhEtAl2021_BiomJ, OhEtAl2021_SIM, Amorim2024} Generally, design-based estimators offer better robustness than model-based ones because they make fewer assumptions about the error mechanism. 

Model-based estimators place an extra model on the error mechanism (i.e., the distribution of map-based $X$ given straight-line $X^*$). Imputation, maximum likelihood estimation, and regression calibration are popular members of this class. \cite{Carroll2006} Replacing missing $X$ values with predictions based on $X^*$ (i.e., imputation) is a promising and popular option because of its ease of implementation and approachability to statisticians and nonstatisticians. So far, multiple imputation has been used to overcome measurement error in binary outcomes or exposures \cite{Cole2006, Edwards2013, Edwards2015, Edwards2020};  in continuous outcomes and covariates \cite{Shepherd2012};  and in a mix of categorical, continuous, or time-to-event variables. \cite{Giganti2020, Han2021, Pelgrims2023, Amorim2024} When assumptions on the error mechanism are correct, model-based estimators can offer better statistical efficiency than design-based ones.

\subsection{Overview}

A new statistical approach is proposed to (i) accurately quantify food access on a large geographic scale from incomplete data and (ii) efficiently model its impact on health with less computational strain. This multiple imputation framework is easily generalizable, as adopting other outcome models, adding other fully observed covariates, and incorporating spatial autocorrelation are very straightforward. Also, food access for all neighborhoods in a study can be predicted despite only validating a subset of them, providing a more accurate and comprehensive estimate of the food access landscape. The rest of the paper is organized as follows. In Section~\ref{sec:meth}, we describe the distance calculations and proposed imputation methods. In Sections~\ref{sec:sims} and \ref{sec:app}, the methods are tested in extensive simulations and applied to data from the Piedmont Triad region in northwestern North Carolina. In Section~\ref{sec:conc}, we discuss our findings and future directions. 

\section{Methods}
\label{sec:meth}

\subsection{Model and data}\label{sec:meth_model}

The neighborhood-level rates of various health outcomes will be modeled in the following way. Specifically, census tracts were adopted as the ``neighborhood'' unit for the analysis in Section~\ref{sec:app} since that was the smallest geographic scale at which the data were available. Let $Y$ be the model outcome, denoting the number of cases in a neighborhood ($Y \in \{0, 1, \dots\}$). This outcome is offset by $Pop$, denoting the neighborhood population ($Pop \in \{Y, Y+1, \dots\})$. The neighborhood's food access will be measured by $X$, denoting the proximity to healthy foods ($X \in 
\mathds{R}^+$) from the population center, or population-weighted centroid. \cite{cb_pop_center} ``Healthy food stores'' are defined in Section~\ref{sec:data_collect} and can be thought of as grocery stores or other fresh food sources. 

Poisson regression will be used to assess the impact of food access (measured by $X$) on the neighborhood-level rate of various  health outcomes ($Y$ per $Pop$), adjusting for other fully observed covariates $\bZ$ (e.g., rural/urban status): 
\begin{align}
    \log\{\E_{\bbeta}(Y|X)\} = \beta_{0} + \beta_{1}X + \bbeta_2^\top\bZ + \log(Pop). \label{mod_prox} 
\end{align}
The adjusted prevalence ratio for food access, $\exp(\beta_{1})$, is of primary interest. An adjusted prevalence ratio of $\exp(\beta_{1}) > 1$ indicates that worse proximity (i.e., farther distances to the neighborhood's nearest healthy food store) is associated with higher prevalence of the health outcome, $\exp(\beta_{1}) = 1$ indicates that proximity had no impact on prevalence of the health outcome, and $\exp(\beta_{1}) < 1$ indicates that worse proximity led to lower prevalence of the health outcome.

\subsection{Distance calculations}\label{sec:meth_dist}

Proximity to healthy foods $X$ is derived from the distances between neighborhood population centers and nearby grocery stores, supermarkets, and other fresh food sources. Suppose data are collected on $N$ neighborhoods and $M$ stores in the study area. If $D_{ij}$ denotes the map-based distance between neighborhood center $i$ ($i \in \{1, \dots, N\}$) and healthy food store $j$ ($j \in \{1, \dots, M\}$), then $X_{i} = \min(D_{i1}, \dots, D_{iM})$ is neighborhood $i$'s map-based proximity to healthy foods. Due to the added computational strain of map-based calculations, the straight-line distances are first calculated for all neighborhoods ($i \in \{1, \dots, n\}$) and stores ($j \in \{1, \dots, M\}$) and then used to narrow down combinations ($i, j$) of the shortest routes to revisit for the map-based distances. 

Straight-line distances $D_{ij}^*$ ($i \in \{1, \dots, N\}$, $j \in \{1, \dots, M\}$) for all neighborhoods and stores are calculated using the Haversine formula, which measures the shortest arc distance between two sets of latitude and longitude coordinates. \cite{Sinnott1984} This formula is implemented in the \texttt{distHaversine} function of the \textit{geosphere} package in R. \cite{Hijmans2022} It can be used to quickly compute the distances between all neighborhoods and all healthy food stores. For the Piedmont Triad analysis in Section~\ref{sec:app}, calculating $387 \times 701 = \num{271287}$ straight-line distances between $N = 387$ neighborhoods and $M = 701$ stores took $0.1$ seconds on a 2020 Macbook Pro (M1 Chip) with 16 gigabytes of memory. Based on straight-line distances $D_{ij}^*$, the error-prone version of $X_{i}$ is defined in parallel as $X_{i}^* = \min(D_{i1}^*, \dots, D_{iM}^*)$.

Select map-based distances $D_{ij}$ ($i \in \{1, \dots, N\}$, $j \in \{1, \dots, M\}$) are calculated using the \texttt{mapdist} function from the \textit{ggmap} package in R to query the shortest driving routes between two sets of latitude and longitude coordinates using the Google Maps API. \cite{KahleWickham2013} In agreement with the API's Terms of Service, individual distances $D_{ij}$ were not saved; only the derived variable $X_{i}$, computed as the minimum over all $j$, was saved. 

We are interested in the distance to the closest store. However, the closest store by the straight-line distance may not be the closest when using the more accurate map-based distance. Thus, when building the Piedmont Triad data, all routes to healthy food stores that were in the lower 20th percentile of straight-line distances for each neighborhood were queried using Google Maps. Across the $N = 387$ neighborhoods, the median distance used as this threshold was $13.4$ miles (interquartile range [IQR] $=[11,22.6]$). The 20th percentile was chosen as a conservative cutoff; it is unlikely that any healthy food stores farther away could have been the neighborhood's nearest. Focusing on routes to the $701 \times 0.2 = 140$ nearest stores reduced computation time. Still, calculating $387 \times 701 \times 0.2 = \num{54257}$ map-based distances took over an hour ($77$ minutes) on the same Macbook Pro, compared to the $0.1$ seconds to calculate all $\num{271287}$ straight-line distances (Supplemental Figure~S2). 

\subsection{Overcoming missing map-based food access data}\label{sec:meth_imputation}

The error-prone food access variable $X^*$ is observed for all neighborhoods. However, due to computational strain, map-based calculations may only done for a subset of $n$ neighborhoods ($n < N$), leading to missing values of the error-free food access variable $X$ for the remaining $N - n$ neighborhoods. The missing variable  will be multiply imputed for these ``unqueried'' neighborhoods. The imputed values are introduced first before discussing the multiple imputation framework.

Missing values of map-based proximity, $X$, will be replaced with random draws from their conditional distributions given their error-prone value, $X^*$, the log-transformed outcome, $\log(Y)$, and additional fully-observed covariates, $\bZ$. There can also be extra fully-observed covariates here (e.g., neighborhood gentrification or car ownership) that are more tailored to capturing the error mechanism through the imputation model rather than contributing to the analysis model. This conditional distribution is assumed to be normal, with mean $= \alpha_{0} + \alpha_{1}X^* + \alpha_{2}\log(Y) + \pmb{\alpha}_3^\top\bZ$ and standard deviation $= \sigma$. Estimated parameters for this distribution, denoted by $\balphahat = (\alphahat_{0}, \alphahat_{1}, \alphahat_{2}, \hat{\pmb{\alpha}}_3, \sigmahat)^{\textrm{T}}$, are obtained by fitting a linear regression ``imputation model'' -- with $X$ as the outcome and $X^*$, $\log(Y)$, and $\bZ$ as the predictors -- to the $n$ ``queried'' neighborhoods without missing data. Note that the analysis model outcome, $\log(Y)$, must be included in the predictor's imputation model for the imputation model to be congenial. \cite{Moons2006, DAgostinoMcGowan2024} In Supplemental Section~S.1, other ways of including the outcome in the imputation model were considered, but using $\log(Y)$ performed best in simulations (Supplemental Figure~S3).

These imputed values are incorporated into a multiple imputation framework to obtain valid standard error estimates for statistical inference about the adjusted prevalence ratio of interest, $\exp(\beta_1)$. Specifically, the following two-step procedure was adopted from prior work for each iteration $b$ ($b \in \{1, \dots, B\}$). \cite{Shepherd2012}
\begin{enumerate}
    \item \textit{Draw imputed values:} Impute missing $X$ with random draws $\breve{X}^{(b)}$ from the conditional distribution parameterized by $\balphahat$. Let $\Xtilde^{(b)}$ denote the imputed variable from iteration $b$, where $\Xtilde^{(b)} = X$ if the map-based measure is non-missing and $\Xtilde^{(b)} =$ $\breve{X}^{(b)}$ otherwise. 
    \item \textit{Fit analysis model to imputed dataset:} Estimate the model of health and food access from Equation~\eqref{mod_prox} with a Poisson regression of $Y$ on $\Xtilde^{(b)}$ and $\bZ$ to obtain parameter and variance estimates, $\bbetahat^{(b)}$ and $\Vhat\left(\bbetahat^{(b)}\right)$, respectively, from iteration $b$.   
\end{enumerate}
For the final analysis, the $B$ sets of estimates are pooled according to Rubin's rules. \cite{rubin2004multiple} This imputation approach is implemented in the \texttt{impPossum} function from the \textit{possum} R package, available on GitHub at \url{https://github.com/sarahlotspeich/possum}.

\section{Simulation Studies}\label{sec:sims}

Using data generated to mimic the Piedmont Triad data analyzed in Section~\ref{sec:app}, we demonstrate how (i) using error-prone, straight-line food access measures can lead to bias but (ii) incorporating more accurate, map-based food access measures for even a small subset of neighborhoods can correct that bias. Further, imputing map-based food access for neighborhoods where it was missing can offer better efficiency than ignoring those neighborhoods. R scripts to reproduce all simulations, tables, and figures are available on GitHub at \url{https://github.com/sarahlotspeich/food_access_imputation}.

\subsection{Setup and data generation}\label{sec:sims_setup}

Samples of $N = 387$ or $2169$ neighborhoods were simulated in the following way. (These sample sizes were chosen to represent the Piedmont Triad region and the entire state of North Carolina, respectively.) First, map-based proximity to the nearest healthy food store (in miles), $X$, was generated from a gamma distribution with shape $= 1$ and scale $= 2.5$. Then, error-prone, straight-line proximity, $X^*$, was constructed following an additive measurement error model as $X^* = X + U$. The error $U$ was generated from a truncated normal distribution with mean $= \mu_U$, standard deviation $= \sigma_U$, lower bound $= -X$, and upper bound $= 0$. (Setting these lower and upper bounds simulates $0 \leq X^* \leq X$, as expected.)  Unless otherwise stated, $\mu_U = -0.7$ and $\sigma_U = 0.8$. Next, the population $Pop$ was generated from a Poisson distribution with mean $ = 4095$ people. Finally, the number of  cases $Y$ was generated from a Poisson distribution with mean $ = Pop \left\{ \exp(-2.2 + 0.01X)\right\}$, leading to 11\% prevalence, on average. A random proportion of $q = 0.1$ neighborhoods were treated as queried (i.e., to have non-missing $X$), while the remaining $1-q = 0.9$ were treated as unqueried (i.e., to have missing $X$). This proportion relates to the validation study size through $n = \nint{Nq}$, where $\nint{\cdot}$ denotes the nearest integer function. The mean population, outcome prevalence, error mean, and error standard deviation were all based on the Piedmont Triad data.

The model in Equation~\eqref{mod_prox} was of interest, which captured the association between the neighborhood-level prevalence of the outcome ($Y / Pop$) and map-based proximity to healthy food stores ($X$) using Poisson regression. For comparison, this model was fit to the unqueried data (i.e., using $X^*$ for all neighborhoods), the fully queried data (i.e., using $X$ for all neighborhoods), and the partially queried data (i.e., using $X$ for a subset of neighborhoods). The \textit{gold standard analysis} used the fully queried data to fit the model using $Y$, $Pop$, and $X$ from all $N$ neighborhoods. The \textit{naive analysis} used the unqueried data to fit the model using $Y$, $Pop$, and $X^*$ (instead of $X$) from all $N$ neighborhoods. The \textit{complete case analysis} used the partially queried data to fit the model using $Y$, $Pop$, and $X$ from only the subset of $n$ neighborhoods. The \textit{imputation analysis} used the partially queried and unqueried data together to fit the model using $Y$, $Pop$, and multiply imputed $\widetilde{X}$ (instead of $X$) from all $N$ neighborhoods $B = 20$ times following the process detailed in Section~\ref{sec:meth_imputation}. 

Empirical relative bias and standard errors for the estimated log prevalence ratio $\betahat_1$ for map-based proximity $X$ are reported for the four analysis approaches to assess their validity for estimation. For imputation, the average standard error estimates are compared to the empirical standard errors to assess their validity for inference. Among unbiased approaches, relative efficiency to the gold standard analysis is reported to compare statistical precision. Relative efficiency is calculated as the ratio of the empirical variances of the gold standard to the other approaches, with ratios closer to one indicating that more efficiency was recovered.

\subsection{Additive errors in straight-line proximity}\label{sec:sims_vary_errors}

Simulation results under increasingly severe additive errors in straight-line proximity to healthy foods $X^*$ are summarized in Table~\ref{tab:vary_sigmaU}. The error magnitude (i.e., severity) was varied by considering different standard deviations $\sigma_U \in \{0.1, 0.2, 0.4, 0.8, 1\}$ for $U$. For all choices of $\sigma_U$, straight-line proximity $X^*$ remained less than or equal to map-based proximity $X$. With $\hat{\sigma}_U = 0.8$ in the Piedmont Triad data (Section~\ref{sec:app:desc_proximity}), our simulations considered more and less severe errors than were in the real data. As expected, the relative bias of the naive analysis grew as $\sigma_U$ increased (from $3\%$ to $6\%$) and persisted regardless of $N$. In all cases, the complete case and imputation analyses offered reduced bias over the naive, particularly under more severe errors. Even with the smallest sample size and most severe errors, imputation was $< 2\%$ biased. 

\begin{table}[ht]
\centering
\caption{Simulation results under increasingly severe additive errors in straight-line proximity to healthy foods, as controlled by the standard deviation $\sigma_U$ of the errors $U$. The mean $\mu_U = -0.7$ of the errors was fixed. \label{tab:vary_sigmaU}
}
\resizebox{\columnwidth}{!}{
\begin{threeparttable}
\centering
\begin{tabular}{rcrcrcrccrcccc}
\toprule
\multicolumn{2}{c}{\textbf{ }} & \multicolumn{2}{c}{\textbf{Gold Standard}} & \multicolumn{2}{c}{\textbf{Naive}} & \multicolumn{3}{c}{\textbf{Complete Case}} & \multicolumn{5}{c}{\textbf{Imputation}} \\
\cmidrule(l{3pt}r{3pt}){3-4} \cmidrule(l{3pt}r{3pt}){5-6} \cmidrule(l{3pt}r{3pt}){7-9} \cmidrule(l{3pt}r{3pt}){10-14}
\textbf{$\pmb{N}$} & \textbf{$\pmb{\sigma_U}$} & \textbf{Bias} & \textbf{ESE} & \textbf{Bias} & \textbf{ESE} & \textbf{Bias} & \textbf{ESE} & \textbf{RE} & \textbf{Bias} & \textbf{ESE} & \textbf{ASE} & \textbf{CP} & \textbf{RE}\\
\midrule
$387$ & $0.10$ & $ 0.003$ & $0.001$ & $0.032$ & $0.001$ & $-0.024$ & $0.003$ & $0.085$ & $-0.003$ & $0.001$ & $0.001$ & $0.954$ & $0.931$\\
 & $0.20$ & $-0.002$ & $0.001$ & $0.029$ & $0.001$ & $-0.005$ & $0.003$ & $0.090$ & $-0.008$ & $0.001$ & $0.001$ & $0.954$ & $0.900$\\
 & $0.40$ & $-0.001$ & $0.001$ & $0.032$ & $0.001$ & $-0.010$ & $0.003$ & $0.075$ & $-0.007$ & $0.001$ & $0.001$ & $0.961$ & $0.794$\\
 & $0.80$ & $-0.003$ & $0.001$ & $0.043$ & $0.001$ & $-0.004$ & $0.003$ & $0.083$ & $-0.015$ & $0.001$ & $0.001$ & $0.968$ & $0.640$\\
 & $1.00$ & $ 0.001$ & $0.001$ & $0.055$ & $0.001$ & $ 0.001$ & $0.003$ & $0.073$ & $-0.014$ & $0.001$ & $0.001$ & $0.967$ & $0.572$\\
\addlinespace
$2169$ & $0.10$ & $ 0.000$ & $0.000$ & $0.029$ & $0.000$ & $-0.004$ & $0.001$ & $0.091$ & $-0.003$ & $0.000$ & $0.000$ & $0.958$ & $0.932$\\
 & $0.20$ & $ 0.000$ & $0.000$ & $0.030$ & $0.000$ & $-0.001$ & $0.001$ & $0.095$ & $-0.002$ & $0.000$ & $0.000$ & $0.957$ & $0.884$\\
 & $0.40$ & $-0.001$ & $0.000$ & $0.031$ & $0.000$ & $-0.005$ & $0.001$ & $0.092$ & $-0.006$ & $0.000$ & $0.000$ & $0.965$ & $0.796$\\
 & $0.80$ & $ 0.001$ & $0.000$ & $0.045$ & $0.000$ & $ 0.003$ & $0.001$ & $0.100$ & $-0.006$ & $0.000$ & $0.001$ & $0.974$ & $0.658$\\
 & $1.00$ & $ 0.001$ & $0.000$ & $0.055$ & $0.000$ & $-0.001$ & $0.001$ & $0.092$ & $-0.007$ & $0.000$ & $0.001$ & $0.979$ & $0.605$\\
\bottomrule
\end{tabular}
\begin{tablenotes}[flushleft]
\item{\em Note:} \simtabnote
\end{tablenotes}
\end{threeparttable}
}
\end{table}

The standard error estimator for imputation performed well, closely approximating the empirical standard error on average. In most settings, empirical coverage probabilities for its 95\% confidence intervals were close to $0.95$. With the worst errors ($\sigma_U = 0.8$ or $1$), the confidence intervals were conservative (i.e., with higher-than-nominal coverage). Since $X$ was simulated from a gamma distribution and the error $U$ added to it from a truncated normal, the normal linear regression imputation model is technically misspecified here, which could explain the inflated coverage. \cite{Hughes2016} Either the Robins and Wang estimator\cite{Robins&Wang2000}  or a bootstrap approach \cite{Bartlett&Hughes2020} might be adopted to obtain narrower confidence intervals. 

Given the small scale of the true $\beta_1 = 0.01$, the standard errors for all approaches were small. Still, the imputation estimator recovered $57-93\%$ of the efficiency of the gold standard analysis, while the complete case analysis only recovered up to $10\%$. With either $N$, the imputation estimator recovered more efficiency with smaller $\sigma_U$, when the relationship between $X$ and $X^*$ was more informative. Similar observations hold when the error mean $\mu_U$ was varied instead (Supplemental Table~S1). Thus, imputation performs well under errors in $X^*$ that underestimated $X$ by a bigger margin (driven by larger $\mu_U$) in addition to errors $X^*$ that were noisier about $X$ (due to larger $\sigma_U$). 

Simulation results with a different proportion $q \in \{0.1, 0.25, 0.5, 0.75\}$ of queried neighborhoods with non-missing $X$ can be found in Table~\ref{tab:vary_q}. All metrics for the naive analysis, which ignores the queried data, were effectively unchanged across these settings. For any sample size $N$, the standard errors and, to a lesser extent, the relative bias of the complete case analysis and imputation estimators decreased as $q$ increased (i.e., when there was less missing data). While the efficiency gains of imputation over the complete case analysis were largest when $q$ was small (relative efficiency $56\%$ larger), they remained sizable (relative efficiency $21-22\%$ larger) even when $q$ was large.

\begin{table}[ht]
\centering
\caption{Simulation results with more neighborhoods queried to obtain map-based proximity to healthy foods, as controlled by the proportion $q = n / N$. The additive errors $U$ in straight-line proximity were generated with mean $\mu_U = -0.7$ and standard deviation $\sigma_U = 0.8$. \label{tab:vary_q}}
\resizebox{\columnwidth}{!}{
\begin{threeparttable}
\begin{tabular}{rcrcrcrccrcccc}
\toprule
\multicolumn{2}{c}{\textbf{ }} & \multicolumn{2}{c}{\textbf{Gold Standard}} & \multicolumn{2}{c}{\textbf{Naive}} & \multicolumn{3}{c}{\textbf{Complete Case}} & \multicolumn{5}{c}{\textbf{Imputation}} \\
\cmidrule(l{3pt}r{3pt}){3-4} \cmidrule(l{3pt}r{3pt}){5-6} \cmidrule(l{3pt}r{3pt}){7-9} \cmidrule(l{3pt}r{3pt}){10-14}
\textbf{$\pmb{N}$} & \textbf{$\pmb{q}$} & \textbf{Bias} & \textbf{ESE} & \textbf{Bias} & \textbf{ESE} & \textbf{Bias} & \textbf{ESE} & \textbf{RE} & \textbf{Bias} & \textbf{ESE} & \textbf{ASE} & \textbf{CP} & \textbf{RE}\\
\midrule
$387$ & $0.10$ & $-0.003$ & $0.001$ & $0.043$ & $0.001$ & $-0.004$ & $0.003$ & $0.083$ & $-0.015$ & $0.001$ & $0.001$ & $0.968$ & $0.640$\\
 & $0.25$ & $ 0.001$ & $0.001$ & $0.048$ & $0.001$ & $-0.018$ & $0.002$ & $0.225$ & $-0.008$ & $0.001$ & $0.001$ & $0.987$ & $0.827$\\
 & $0.50$ & $ 0.001$ & $0.001$ & $0.047$ & $0.001$ & $ 0.004$ & $0.001$ & $0.494$ & $-0.002$ & $0.001$ & $0.001$ & $0.983$ & $0.914$\\
 & $0.75$ & $-0.002$ & $0.001$ & $0.044$ & $0.001$ & $-0.001$ & $0.001$ & $0.751$ & $-0.004$ & $0.001$ & $0.001$ & $0.978$ & $0.975$\\
\addlinespace
$2169$ & $0.10$ & $ 0.001$ & $0.000$ & $0.045$ & $0.000$ & $0.003$ & $0.001$ & $0.100$ & $-0.006$ & $0.000$ & $0.001$ & $0.974$ & $0.658$\\
 & $0.25$ & $-0.001$ & $0.000$ & $0.045$ & $0.000$ & $0.002$ & $0.001$ & $0.249$ & $-0.006$ & $0.000$ & $0.001$ & $0.983$ & $0.843$\\
 & $0.50$ & $ 0.000$ & $0.000$ & $0.045$ & $0.000$ & $0.001$ & $0.001$ & $0.501$ & $-0.003$ & $0.000$ & $0.000$ & $0.976$ & $0.957$\\
 & $0.75$ & $ 0.000$ & $0.000$ & $0.046$ & $0.000$ & $0.002$ & $0.000$ & $0.760$ & $-0.001$ & $0.000$ & $0.000$ & $0.965$ & $0.973$\\
\bottomrule
\end{tabular}
\begin{tablenotes}[flushleft]
\item{\em Note:} \simtabnote
\end{tablenotes}
\end{threeparttable}
}
\end{table}

Table~\ref{tab:vary_prev} summarizes simulations where the disease prevalence and prevalence ratio for map-based food access $X$ were varied. Average disease prevalences (hereafter, simply ``prevalences'') of $7\%$, $11\%$, and $33\%$ were considered by varying $\beta_0 \in \{-2.7, -2.2, -1.1\}$, respectively, and prevalence ratios of $\exp(\beta_1) \in \{0.95, 0.99, 1, 1.01, 1.05\}$ were included. At the null (i.e., $\beta_1 = 0$), all methods were unbiased, regardless of the  prevalence. Beyond that setting, the relative bias of the naive analysis increased as $\beta_1$ got farther from zero but was comparable across prevalences. The complete case and imputation estimators were both virtually unbiased ($<2\%$), and the confidence intervals for the imputation estimator achieved proper coverage in all settings. The complete case analysis only recovered $7-10\%$ of the efficiency of the gold standard. Meanwhile, imputation recovered up to $72\%$ at the null but less as $\beta_1$ moved away from zero. 

\begin{table}[ht]
\centering
\caption{Simulation results under higher disease prevalence and varied prevalence ratios for map-based proximity to healthy foods, as controlled by the coefficients $\beta_0$ (\textbf{Prev.} $=\exp(\beta_0)$) and $\beta_1$ (\textbf{PR} $=\exp(\beta_1)$), respectively. \label{tab:vary_prev}} 
\resizebox{\columnwidth}{!}{
\begin{threeparttable}
\begin{tabular}{rcrcrcrccrcccc}
\toprule
\multicolumn{2}{c}{\textbf{ }} & \multicolumn{2}{c}{\textbf{Gold Standard}} & \multicolumn{2}{c}{\textbf{Naive}} & \multicolumn{3}{c}{\textbf{Complete Case}} & \multicolumn{5}{c}{\textbf{Imputation}} \\
\cmidrule(l{3pt}r{3pt}){3-4} \cmidrule(l{3pt}r{3pt}){5-6} \cmidrule(l{3pt}r{3pt}){7-9} \cmidrule(l{3pt}r{3pt}){10-14}
\textbf{Prev.} & \textbf{PR} & \textbf{Bias} & \textbf{ESE} & \textbf{Bias} & \textbf{ESE} & \textbf{Bias} & \textbf{ESE} & \textbf{RE} & \textbf{Bias} & \textbf{ESE} & \textbf{ASE} & \textbf{CP} & \textbf{RE}\\
\midrule
$0.07$ & $0.95$ & $ 0.001$ & $0.002$ & $0.058$ & $0.002$ & $-0.001$ & $0.005$ & $0.096$ & $-0.002$ & $0.002$ & $0.003$ & $0.968$ & $0.389$\\
 & $0.99$ & $-0.004$ & $0.001$ & $0.045$ & $0.001$ & $-0.006$ & $0.004$ & $0.086$ & $-0.015$ & $0.002$ & $0.002$ & $0.977$ & $0.680$\\
 & $1.00$ & $ 0.000$ & $0.001$ & $0.000$ & $0.001$ & $ 0.000$ & $0.004$ & $0.079$ & $ 0.000$ & $0.001$ & $0.002$ & $0.974$ & $0.720$\\
 & $1.01$ & $ 0.008$ & $0.001$ & $0.052$ & $0.001$ & $ 0.020$ & $0.004$ & $0.080$ & $ 0.002$ & $0.001$ & $0.002$ & $0.974$ & $0.692$\\
 & $1.05$ & $-0.001$ & $0.001$ & $0.039$ & $0.001$ & $-0.005$ & $0.004$ & $0.070$ & $-0.012$ & $0.002$ & $0.002$ & $0.933$ & $0.228$\\
\addlinespace
$0.11$ & $0.95$ & $ 0.000$ & $0.001$ & $0.056$ & $0.001$ & $ 0.004$ & $0.004$ & $0.085$ & $-0.002$ & $0.002$ & $0.003$ & $0.952$ & $0.281$\\
 & $0.99$ & $ 0.007$ & $0.001$ & $0.056$ & $0.001$ & $ 0.002$ & $0.003$ & $0.095$ & $-0.002$ & $0.001$ & $0.001$ & $0.972$ & $0.648$\\
 & $1.00$ & $ 0.000$ & $0.001$ & $0.000$ & $0.001$ & $ 0.000$ & $0.003$ & $0.080$ & $ 0.000$ & $0.001$ & $0.001$ & $0.975$ & $0.699$\\
 & $1.01$ & $-0.004$ & $0.001$ & $0.041$ & $0.001$ & $-0.005$ & $0.003$ & $0.076$ & $-0.017$ & $0.001$ & $0.001$ & $0.964$ & $0.597$\\
 & $1.05$ & $ 0.000$ & $0.001$ & $0.040$ & $0.001$ & $ 0.001$ & $0.003$ & $0.070$ & $-0.011$ & $0.002$ & $0.002$ & $0.943$ & $0.173$\\
\addlinespace
$0.33$ & $0.95$ & $0.000$ & $0.001$ & $0.055$ & $0.001$ & $-0.003$ & $0.002$ & $0.082$ & $-0.009$ & $0.002$ & $0.002$ & $0.951$ & $0.151$\\
 & $0.99$ & $0.001$ & $0.001$ & $0.050$ & $0.001$ & $-0.006$ & $0.002$ & $0.079$ & $-0.013$ & $0.001$ & $0.001$ & $0.968$ & $0.515$\\
 & $1.00$ & $0.000$ & $0.001$ & $0.000$ & $0.001$ & $ 0.000$ & $0.002$ & $0.080$ & $ 0.000$ & $0.001$ & $0.001$ & $0.968$ & $0.697$\\
 & $1.01$ & $0.002$ & $0.001$ & $0.048$ & $0.001$ & $ 0.001$ & $0.002$ & $0.077$ & $-0.015$ & $0.001$ & $0.001$ & $0.960$ & $0.531$\\
 & $1.05$ & $0.000$ & $0.000$ & $0.040$ & $0.001$ & $-0.001$ & $0.002$ & $0.076$ & $-0.011$ & $0.002$ & $0.002$ & $0.946$ & $0.096$\\
\bottomrule
\end{tabular}
\begin{tablenotes}[flushleft]
\item{\em Note:} \simtabnote
\end{tablenotes}
\end{threeparttable}
}
\end{table}

\subsection{Multiplicative errors in straight-line proximity} \label{sec:sims_mult}

Error-prone $X^*$ was also simulated from a multiplicative error model such that $X^* = WX$. Random errors $W$ were generated from a truncated Normal distribution with mean $= \mu_W$, standard deviation $= \sigma_W$, lower bound $= 0$, and upper bound $= 1$. (Setting the maximum to $1$ simulated straight-line proximity to be strictly less than or equal to map-based, as expected.) Fixing $\mu_W = 0.7$, different standard deviations $\sigma_W \in \{0.1, 0.15, 0.2\}$ were considered for $W$, where smaller values were expected to yield less severe errors in $X^*$. However, since the error mean of $0.7$ was close to the upper bound of $1$, larger values of $\sigma_W$ could actually lead to some more severe errors (with $W$ closer to $0$) \textit{and} some less severe errors (with $W$ closer to $1$). With $\hat{\sigma}_W = 0.15$ in the Piedmont Triad data, settings with more and less error variability were again considered. All other variables were simulated following Section~\ref{sec:sims_setup}.

Simulation results under varied standard deviations for the multiplicative error  in straight-line proximity to healthy foods are described in Table~\ref{tab:mult_errors}. The naive analysis was considerably more biased in the multiplicative error settings considered than under the additive ones in Section~\ref{sec:sims_vary_errors} (as high as $38\%$ versus $6\%$). The complete case analysis offered low bias ($< 3\%$) but low efficiency ($10\%$ or less recovered) in all settings. The imputation estimator offered similarly low bias ($< 2\%$) with much higher efficiency ($34-62\%$ recovered). Its standard error estimator and confidence intervals remained valid in all settings. 

\begin{table}[ht]
\centering
\caption{Simulation results under varied multiplicative errors in straight-line proximity to healthy foods, as controlled by the standard deviation $\sigma_W$ of the errors $W$. The mean $\mu_W = 0.7$ of the errors was fixed. \label{tab:mult_errors}}
\resizebox{\columnwidth}{!}{
\begin{threeparttable}
\begin{tabular}{rcrcrcrccrcccc}
\toprule
\multicolumn{2}{c}{\textbf{ }} & \multicolumn{2}{c}{\textbf{Gold Standard}} & \multicolumn{2}{c}{\textbf{Naive}} & \multicolumn{3}{c}{\textbf{Complete Case}} & \multicolumn{5}{c}{\textbf{Imputation}} \\
\cmidrule(l{3pt}r{3pt}){3-4} \cmidrule(l{3pt}r{3pt}){5-6} \cmidrule(l{3pt}r{3pt}){7-9} \cmidrule(l{3pt}r{3pt}){10-14}
\textbf{$\pmb{N}$} & \textbf{$\pmb{\sigma_W}$} & \textbf{Bias} & \textbf{ESE} & \textbf{Bias} & \textbf{ESE} & \textbf{Bias} & \textbf{ESE} & \textbf{RE} & \textbf{Bias} & \textbf{ESE} & \textbf{ASE} & \textbf{CP} & \textbf{RE}\\
\midrule
$387$ & $0.10$ & $0.000$ & $0.001$ & $0.375$ & $0.001$ & $-0.022$ & $0.003$ & $0.083$ & $-0.006$ & $0.001$ & $0.001$ & $0.936$ & $0.619$\\
 & $0.15$ & $0.001$ & $0.001$ & $0.335$ & $0.001$ & $ 0.013$ & $0.003$ & $0.086$ & $-0.007$ & $0.001$ & $0.001$ & $0.937$ & $0.466$\\
 & $0.20$ & $0.001$ & $0.001$ & $0.308$ & $0.001$ & $-0.022$ & $0.003$ & $0.080$ & $-0.014$ & $0.002$ & $0.002$ & $0.937$ & $0.363$\\
\addlinespace
$2169$ & $0.10$ & $ 0.000$ & $0.000$ & $0.374$ & $0.001$ & $-0.002$ & $0.001$ & $0.093$ & $-0.004$ & $0.001$ & $0.001$ & $0.944$ & $0.555$\\
 & $0.15$ & $ 0.003$ & $0.000$ & $0.335$ & $0.001$ & $ 0.006$ & $0.001$ & $0.098$ & $-0.005$ & $0.001$ & $0.001$ & $0.934$ & $0.384$\\
 & $0.20$ & $-0.001$ & $0.000$ & $0.306$ & $0.001$ & $ 0.004$ & $0.001$ & $0.082$ & $-0.015$ & $0.001$ & $0.001$ & $0.950$ & $0.340$\\
\bottomrule
\end{tabular}
\begin{tablenotes}[flushleft]
\item{\em Note:} \simtabnote
\end{tablenotes}
\end{threeparttable}
}
\end{table}

As with additive errors, the imputation estimator recovered more efficiency in the settings with smaller $\sigma_W$. However, whereas larger $\sigma_U$ led to larger bias for the naive analysis with additive errors, larger $\sigma_W$ led to smaller bias for the naive analysis with multiplicative errors. Intuitively, this result is likely because the mean $\mu_W = 0.7$ was already close to the upper bound of $1$. Thus, while increasing $\sigma_W$ pushed some simulated errors $W$ toward $0$ (more severe), it also led to more errors near $1$ (less severe). In additional simulations, we varied the multiplicative error mean $\mu_W \in \{0.3, 0.5, 0.7\}$ but held the standard deviation $\sigma_W = 0.15$ fixed. We saw that pushing the mean closer to the lower bound of $0$ instead led to larger bias for the naive analysis, as we incurred more small values of $W$ and thus more severe errors (Supplemental Table~S2).

\section{Application to the Piedmont Triad, North Carolina}
\label{sec:app}

The Piedmont Triad, located in the northwestern region of North Carolina, comprises twelve counties (Supplemental Figure~S4). Beyond two medium-sized cities (Greensboro and Winston--Salem), much of the area is rural, with a median population density for neighborhoods (census tracts) across the region of $843$ people per square mile (minimum $= 33$, maximum $= 6681$). The region includes one neighborhood with zero population, which contains only the Piedmont Triad International Airport and was excluded from all analyses. The remaining $N = 387$ neighborhoods in the Piedmont Triad are described in Supplemental Table~S3. The number of neighborhoods per county varied from as few as six in the rural counties to as many as $118$ in the urban ones. According to the 2010 U.S. Census, the median population per neighborhood was $4095$ people (IQR $=[3901,5282]$). Land area per neighborhood was $5$ square miles, on average (IQR $=[2,19]$). 

The socioeconomic factors vary substantially between neighborhoods in this region. A brief socioeconomic profile based on the 2015 American Community Survey (ACS) is illustrated in Supplemental Figure~S5. The median household income was \$53,750, on average, but was as low as \$13,447 and as high as \$176,875. The percentage of neighborhood households with income below the poverty line varies widely and ranges from $0\%$ to $73\%$. Most neighborhoods had low proportions of households receiving the Supplemental Nutrition Assistance Program (SNAP) (median $= 13\%$). Still, there were some neighborhoods with as many as $75\%$ who were receiving this type of government assistance. Most workers ($\geq 45\%$) in nearly all neighborhoods drive alone to work; still, a neighborhood where $97\%$ of workers drive alone (the maximum) would be very different than one where $45\%$ do (the minimum). The proportion of people with health insurance (public or private) varied some (between $65\%$ and $100\%$), although not as much as other variables like the proportion of people $25$ years or older who completed at least some college (between $26\%$ and $96\%$) and the others mentioned thus far.

The average White and Black populations (as percents) per neighborhood in the Piedmont Triad were similar to the state averages of (78\% versus 75\% and 14\% versus 15\%). Like the rest of the state, the Piedmont Triad had extremely low percentages of Asian, American Indian or Alaska Native, or Native Hawaiian and Other Pacific Islander populations (all $\approx 0\%$, on average). Still, there was wide variability between neighborhoods (Supplemental Figure~S6). In particular, some neighborhoods were more than 75\% Black, and these neighborhoods were located primarily in either the urban areas surrounding the cities of Winston-Salem and Greensboro or in the rural county in the northeastern corner of the Piedmont Triad. While the average percent Asian population across the region was small, there were some neighborhoods with more than 20\% Asian population, as well.

\subsection{Motivation for analysis}

\begin{figure}[ht]
     \centering
     \includegraphics[width=\textwidth]{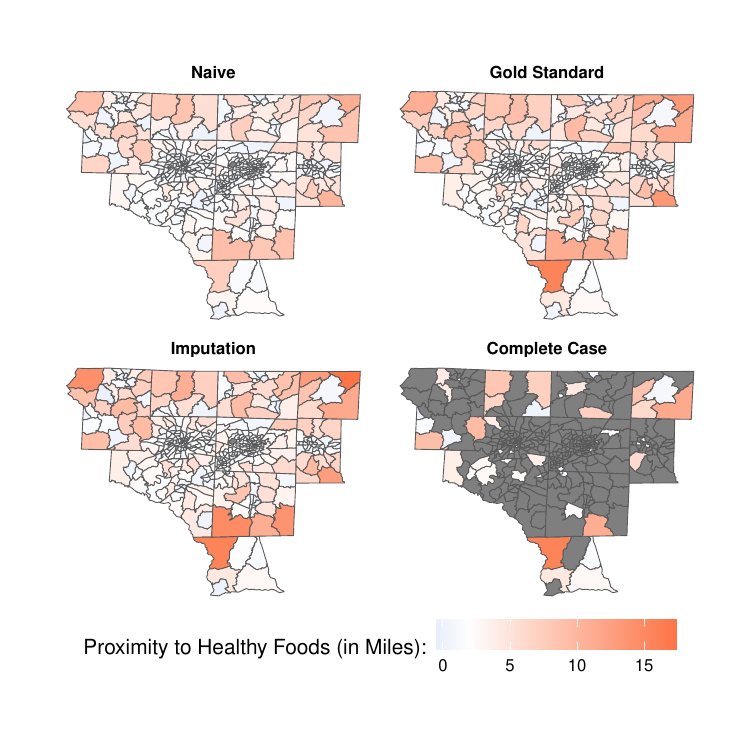}
    \caption{Choropleth map of food access, as measured by proximity to healthy foods for each neighborhood (census tract) in the Piedmont Triad, North Carolina, according to the data used for the naive, gold standard, imputation, and complete case analyses. One census tract was excluded because it had a population of zero. \label{fig:map_food}}
\end{figure}

Preliminary data using straight-line proximity to healthy foods suggested that many neighborhoods in the Piedmont Triad, NC, have few options close to home (Figure~\ref{fig:map_food}). We knew that the actual number of neighborhoods with poor access to healthy foods was likely even higher since map-based proximity had to be as bad as or worse than straight-line proximity. This concern was echoed by data reported in the U.S. Department of Agriculture's Food Research Atlas, \cite{USDA_FoodAtlas} wherein $46\%$ and $67\%$ of neighborhoods in the Piedmont Triad were found to have high proportions of people living more than $1$ mile and $0.5$ mile, respectively, from their nearest supermarket. Many neighborhoods without access were in Forsyth and Guilford County, putting these counties among the least accessible in the state. Interestingly, Forsyth and Guilford are both relatively urban counties, whereas one might expect lower access in more expansive, rural areas. However, beyond these counties' largest cities (Winston-Salem and Greensboro, respectively), there are many rural neighborhoods that could help explain this result.

National studies in the U.S. have previously found links between diabetes and obesity and the availability of healthy foods. Ahern et al. found that lower county-level prevalence of type 2 diabetes was associated with having more ``healthy'' food options (full-service restaurants, grocery stores, direct farm sales). \cite{Ahern2011} Kanchi et al. found that higher relative densities of supermarkets to other retail food outlets were associated with lessened risk of incident type 2 diabetes in a national cohort study of U.S. veterans. \cite{Kanchi2021} Ahern et al. also found full-service restaurants and direct farm sales to be associated with lower county-level prevalence of obesity, while grocery stores were associated with higher rates. \cite{Ahern2011} However, other studies have found conflicting evidence on these connections between food access and health, \cite{White2007} 
so it is critical to re-investigate these relationships for the Piedmont Triad alone. 

\subsection{Data collection}\label{sec:data_collect}

Publicly available data were combined from three sources for the analysis using open-source tools in R. \cite{R} The \textit{tidycensus} package was used to extract ACS data and create maps. \cite{tidycensus} Links to the datasets and all R code are available on GitHub at \url{https://github.com/sarahlotspeich/food_access_imputation}.

Diabetes and obesity prevalence estimates were taken from the U.S. Centers for Disease Control and Prevention's 2022 PLACES dataset. \cite{data_places} Like other administrative datasets, PLACES contains small area estimates (SAEs) of prevalence, which were obtained via statistical modeling using individual- and area-level information (e.g., age, race, education, poverty) as predictors. \cite{PLACES_Methods} The PLACES data were available at the census tract level, which was adopted as the ``neighborhood'' unit for analysis. 

Store locations and types were obtained from the U.S. Department of Agriculture's 2022 Historical SNAP Retail Locator Data to calculate neighborhoods' proximity to healthy foods. \cite{data_snap}  The analysis used $M = 701$ NC farmers' markets, grocery stores (any size), specialties (baker, fruits/vegetables, meat/poultry, seafood), and supermarkets/stores as ``healthy food stores.'' A map of stores by type can be found in Supplemental Figure~S7. To adjust for the urbanicity of the census tracts in analyses, the 2010 Rural-Urban Commuting Area (RUCA) codes were also extracted from the U.S. Department of Agriculture. \cite{data_ruca}

With a moderate number of neighborhoods in the Piedmont Triad ($N = 387 $ census tracts), we were able to establish the gold standard here, which is not always possible in practice. Using Google Maps and the Haversine formula for distance calculations, we collected map-based and straight-line proximity, respectively, to healthy foods for the entire region. Then, map-based proximity $X$ from these fully queried data was used to fit the gold standard analysis. The naive analysis used straight-line proximity $X^*$ from the unqueried data instead. 

For the complete case and imputation analyses, an artificial partially queried dataset was also constructed, wherein only $n = 48$ census tracts ($q = 0.12$) had map-based $X$ available. To ensure the geographic diversity of this queried subset, four neighborhoods were chosen from each county at random to be treated as queried and have map-based proximity $X$ available (Supplemental Figure~S8). This design targets equal representation from counties made up of (i) a large number of small, urban census tracts and (ii) a small number of large, rural census tracts. The former can be found in the center of the Piedmont Triad, where large cities like Winston-Salem and Greensboro are located, while the latter are found along the border of the region. Many other sampling designs are possible, and analysis-specific, targeted strategies are a promising direction for future work. 

\subsection{Health in the Piedmont Triad}

\begin{figure}[ht]
\begin{center}
\includegraphics[width=\textwidth]{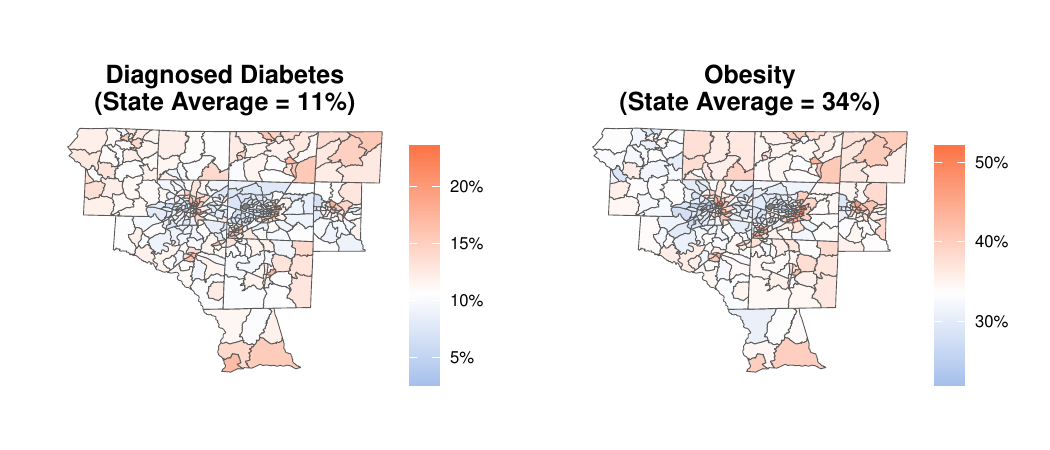}
\end{center}
\caption{Choropleth maps of the crude prevalence of diagnosed diabetes and obesity for census tracts in the Piedmont Triad, North Carolina. The gradient for each map is centered at the state average (median). Data were taken from the 2022 PLACES dataset. \cite{data_places} One census tract was excluded because it had a population of zero.\label{fig:map_health}}
\end{figure}

There was between-neighborhood variability in the prevalences of diagnosed diabetes and obesity outcomes across the Piedmont Triad (Figure~\ref{fig:map_health}). Contrasts were seen between (i) rural and urban neighborhoods but also (ii) neighborhoods surrounding the same major city. Overall, the regional average prevalences for diagnosed diabetes (median $= 0.11$, IQR $=[0.09,0.13]$) and obesity (median $= 0.34$, IQR $=[0.31,0.38]$) were comparable to the state average prevalences ($0.1$ and $0.34$, respectively). Most larger, rural neighborhoods had prevalences of diabetes and obesity above the state averages. In contrast, many smaller urban neighborhoods saw lower burdens of these diseases. 

When zooming in on Forsyth or Guilford County, there seem to exist divisions between neighborhoods with low and high prevalences on opposite sides of their major cities' downtown areas (Supplemental Figure~S9). Socioeconomic differences could potentially drive these divisions, further highlighting the complexity of accounting for the complete picture when analyzing social determinants of health. Toward capturing some of these differences, we incorporated an indicator (denoted by $Z$) of neighborhoods being ``metropolitan'' areas (i.e.,  having RUCA codes $1$, $2$, or $3$). By this definition, there were $326$ ($84\%$) metropolitan neighborhoods in the Piedmont Triad. The analyses that follow (i) incorporated metropolitan status into imputing proximity to healthy foods (Section~\ref{piedmont_imp}) and (ii) adjusted for metropolitan status directly and allowed for an interaction with proximity to healthy foods in the disease prevalence models (Section~\ref{piedmont_models}).

\subsection{Proximity to healthy foods in the Piedmont Triad} \label{sec:app:desc_proximity}

\begin{figure}[ht]
\begin{center}
\includegraphics[width=\textwidth]{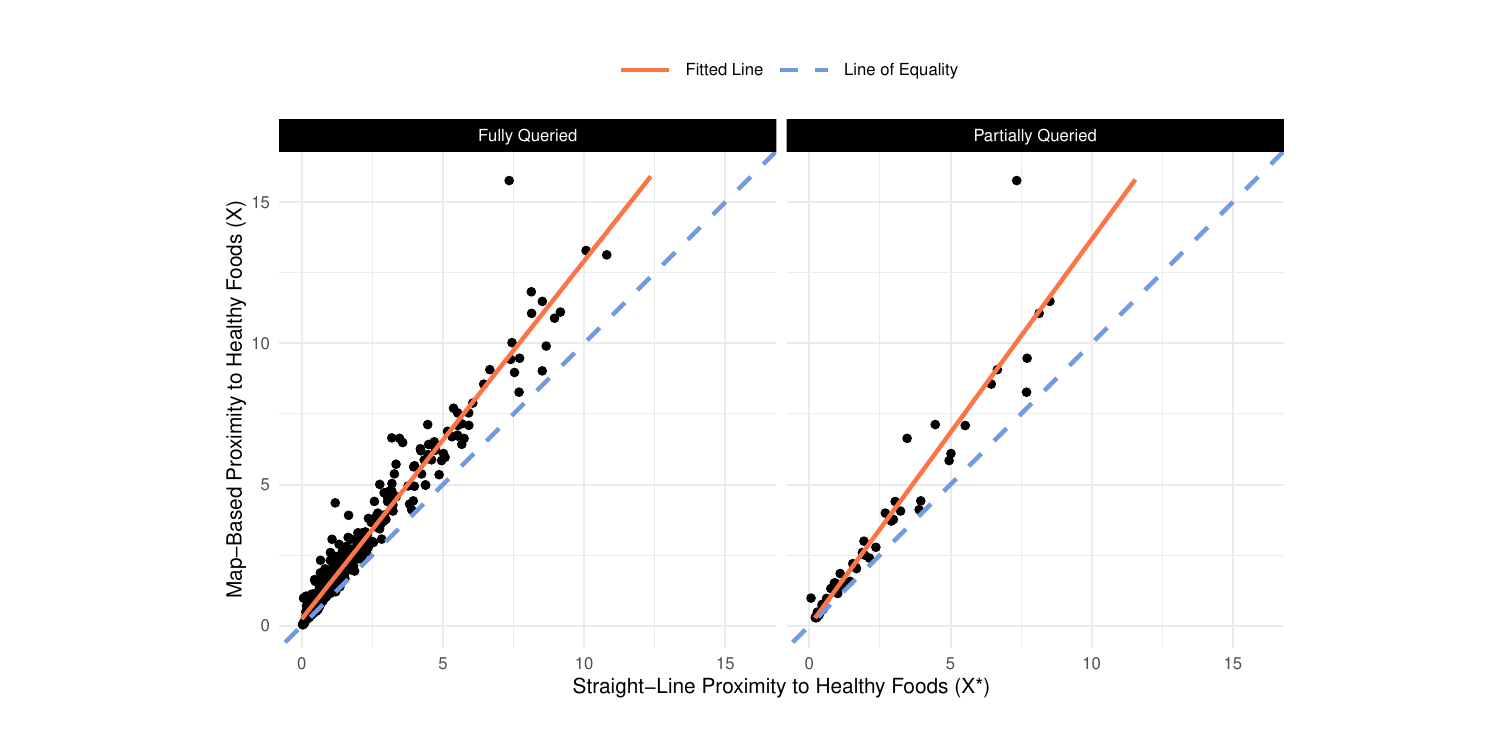}
\end{center}
\caption{Scatter plot of straight-line versus map-based proximity to healthy food store for neighborhoods (census tracts) in the Piedmont Triad, North Carolina using the fully queried data ($N = 387$) or the partially queried data ($n = 48$). The solid line follows the fitted least-squares linear regression between $X$ and $X^*$, while the dashed line denotes the hypothetical $X = X^*$ if there had been no errors in $X^*$ (i.e., the line of equality). 
\label{fig:scatter_proximity}}
\end{figure}

From the fully queried data ($N = 387$ neighborhoods), the following could be learned about straight-line and map-based food access, as well as the relationship between them. Across all neighborhoods in the Piedmont Triad, proximity to healthy foods based on map-based distances was farther, on average, than the same measure based on straight-line distances, as expected (median proximity $= 1.51$ versus $1.01$ miles). The variability of map-based proximity was larger than that of straight-line proximity, also (IQR $= [0.89,2.99]$ versus $[0.57,2.12]$). Still, the two proximity measures were very highly correlated ($R = 0.97$), and the relationship between them was linear (Figure~\ref{fig:scatter_proximity}). Importantly, the relationship between $X$ and $X^*$ in the fully queried and partially queried datasets was the same, with both datasets agreeing that map-based proximity $X$ was worse (i.e., larger) than straight-line proximity $X^*$, particularly at higher values of $X^*$. Thus, the partially queried data used to build the imputation model represented the fully queried (gold standard) data well. The same can be said for this relationship after further breaking down by metropolitan versus non-metropolitan neighborhoods (Supplemental Figure~S10), except that among the non-metropolitan neighborhoods, the partially queried relationship was slightly steeper (indicating more severe errors expected).

When comparing food access between neighborhoods, the straight-line and map-based measures depicted similar trends across the Piedmont Triad, with better proximity (shorter distances) around urban areas and worse proximity (larger distances) in the rural surrounds (Figure~\ref{fig:map_food}). Some rural tracts with poor straight-line proximity had even worse access to healthy foods, according to the map-based measure. For example, the nearest healthy food store to the population center of Census Tract 9301 within Caswell County was $10.8$ miles away according to the straight-line distances (the worst in the Piedmont Triad). Using map-based distances, proximity to healthy foods in this neighborhood was $13.1$ miles (the third-worst instead). There were also instances where straight-line proximity was pretty poor, but map-based proximity uncovered noticeably worse access to healthy foods. For example, Census Tract 9603 within Montgomery County had straight-line proximity of $7.3$ miles (among the 15 worst in the Piedmont Triad) but map-based proximity of $15.8$ miles (the worst in the region). 

The magnitude of the errors in straight-line proximity was also calculated using the fully queried data. First, an additive error mechanism was considered. On average, straight-line proximity $X^*$ underestimated map-based proximity $X$ by $0.7$ miles (i.e., $\hat{\mu}_U = -0.7$). Most neighborhoods ($77\%$) had straight-line proximity $X^*$ that underestimated their map-based proximity $X$ by no more than one mile.  However, the distribution of additive errors across the Piedmont Triad was left-skewed (Supplemental Figure~S11), with the largest additive error of $U=-8.42$. This skewness can also be seen when comparing the mean and median magnitude of the errors, $-0.7$ and $-0.5$, respectively, as the former is influenced more by the larger errors in the tail of the distribution. The standard deviation of the additive errors $\hat{\sigma}_U = 0.8$ was one of the most severe settings considered in the simulations (Section~\ref{sec:sims_vary_errors}). 

Next, a multiplicative error mechanism was considered. On average, straight-line proximity underestimated map-based proximity by $30\%$ (i.e., $\hat{\mu}_W = 0.70$). The distribution of the multiplicative errors was much more symmetric than that of the additive ones (Supplemental Figure~S11). There were still some neighborhoods with much more extreme errors than others; interestingly, they differed from the neighborhoods with the worst additive ones. The most severe multiplicative error ($W=0.063$) was seen for Census Tract 315.04 within Randolph County, which was among the neighborhoods with additive errors under one mile in magnitude ($U=-0.92$). The standard deviation of the multiplicative errors $\hat{\sigma}_W = 0.15$ was chosen as the moderate setting for simulations (Section~\ref{sec:sims_mult}). 

\subsection{Imputation models for straight-line proximity}\label{piedmont_imp}

For each health outcome, the imputation of map-based proximity to healthy foods $X$ began by fitting the linear regression imputation model described in Section~\ref{sec:meth_imputation}. Specifically, $X$ was modeled conditionally on straight-line proximity $X^*$, metropolitan status $Z$, and the log-transformed outcome $\log(Y)$. An added interaction between $X^*$ and $Z$ allowed for different error mechanisms in metropolitan versus non-metropolitan neighborhoods, as is expected, for example, due to infrastructure differences. For diagnosed diabetes, $\hat{\mu} = -0.08 + 1.67X^* + 0.88 Z - 0.42 X^* \times Z - 0.09 \log(Y) \text{ and } \hat{\sigma} = 0.96$. For obesity, $\hat{\mu} = 1.11 + 1.67X^* + 0.90 Z - 0.42 X^* \times Z - 0.24 \log(Y) \text{ and } \hat{\sigma} = 0.95$. For simplicity and given the high adjusted $R^2 \approx 0.93$, other covariates (e.g., road density) were not included in the imputation models, but doing so would be straightforward. 

Using these estimates, each neighborhood missing $X$ in the partially queried dataset was multiply imputed $B = 20$ times for the imputation analyses. Imputing the missing $X$ values for the unqueried neighborhoods allowed them to contribute to the disease models in Section~\ref{piedmont_models}. With only minor modifications, missing $X$ values could also be imputed to supplement the subset of queried data and create a more accurate and complete description of the landscape of food access in the Piedmont Triad than the naive and complete case analyses, respectively. In Figure~\ref{fig:map_food}, the map of proximity to healthy foods based on a single deterministic imputation, not including $\log(Y)$ or resampling, is (i) far more comprehensive than the one based on the complete case (ignoring data for $339$ of $387$ neighborhoods) and (ii) more similar to the gold standard than the one based on the naive (using error-prone straight-line data).

\subsection{Modeling the connection between proximity and disease}\label{piedmont_models}

Finally, the connections between proximity to healthy foods and the prevalence of each health outcome across neighborhoods (census tracts) in the Piedmont Triad were modeled using Poisson regression. In contrast to the simulations, neighborhoods closer together in space could be more similar than neighborhoods farther away. To evaluate this potential spatial correlation, Moran's I was calculated for both diseases of interest using the \textit{spdep} package in R. \cite{spdep} Specifically, Moran's I was calculated for the residuals $e_i^* = Y_i - \widehat{Y}_i^*$ from the naive analysis, where $Y_i$ and $\widehat{Y}_i^* = \exp(\hat{\beta}_0^*+\hat{\beta}_1^*X_i+\hat{\beta}_2^*Z_i+\hat{\beta}_3^*X_i \times Z_i)Pop_i$ denote the observed and predicted number of cases in neighborhood $i$, respectively. The parameters $(\hat{\beta}_0^*,\hat{\beta}_1^*,\hat{\beta}_2^*,\hat{\beta}_3^*)$ were taken from the naive models, which would have fully available data in practice and should be highly correlated with the gold standard ones. Both diagnosed diabetes ($I = 0.06$, $p = 0.02$) and obesity ($I = 0.05$, $p = 0.03$) had spatial autocorrelation that would be considered statistically significant at the $\alpha = 0.05$ significance level. 

A conditional autoregressive (CAR) model was used to capture the associations between proximity to healthy foods and health outcomes (adjusting for metropolitan status) while allowing for spatial autocorrelation between neighborhoods that share a border. This model was defined as a mixed-effects Poisson regression with a random intercept for each neighborhood based on its bordering neighborhoods. Thus, instead of Equation~\eqref{mod_prox}, 
\begin{align}
    \log\{\E_{\bbeta}(Y|X, Z, r)\} = \beta_{0} + \beta_{1}X + \beta_{2}Z + \beta_{3}X \times Z + r + \log(Pop) \label{mod_prox_mixed} 
\end{align}
was of interest in the Piedmont Triad data, where $r$ denotes a neighborhood-specific random intercept. The vector of all neighborhoods' random intercepts $\pmb{r} = (r_1, \dots, r_{397})^\top$ was assumed to follow a multivariate normal distribution with mean vector $\pmb{0}$ and covariance matrix $\pmb{\Sigma}$ defined to allow bordering neighborhoods to be correlated. See Waller and Gotway\cite{WallerGotway2004} for more technical details on the CAR model. While the random intercepts $\pmb{r}$ could be correlated with each other, they were independent of the access variables $(X_1, \dots, X_{397})^\top$ and metropolitan indicators $(Z_1, \dots, Z_{397})^\top$. Using the \texttt{fitme} function from the \textit{spaMM} package in R, \cite{spaMM} Equation~\eqref{mod_prox_mixed} was estimated following the gold standard, naive, complete case, and imputation analysis approaches. For reference, the non-spatial models estimating Equation~\eqref{mod_prox} were also fit to examine the impact of ignoring the spatial autocorrelation.

Additional simulations comparing the gold standard, naive, complete case, and imputation approaches for the mixed-effects model are included in Supplemental Section~S.2. Under different amounts of variability in the spatial random effect, the imputation estimator exhibited low bias ($<2\%$), coverage probabilities close to the nominal $95\%$, and much higher relative efficiency to the gold standard ($0.64 - 0.67$) than the complete case analysis ($0.08 - 0.10$). More results from these simulations can be found in Supplemental Table~S4.

\subsection{Findings from the Piedmont Triad}

\begin{figure}[ht]
\begin{center}
\includegraphics[width=\textwidth]{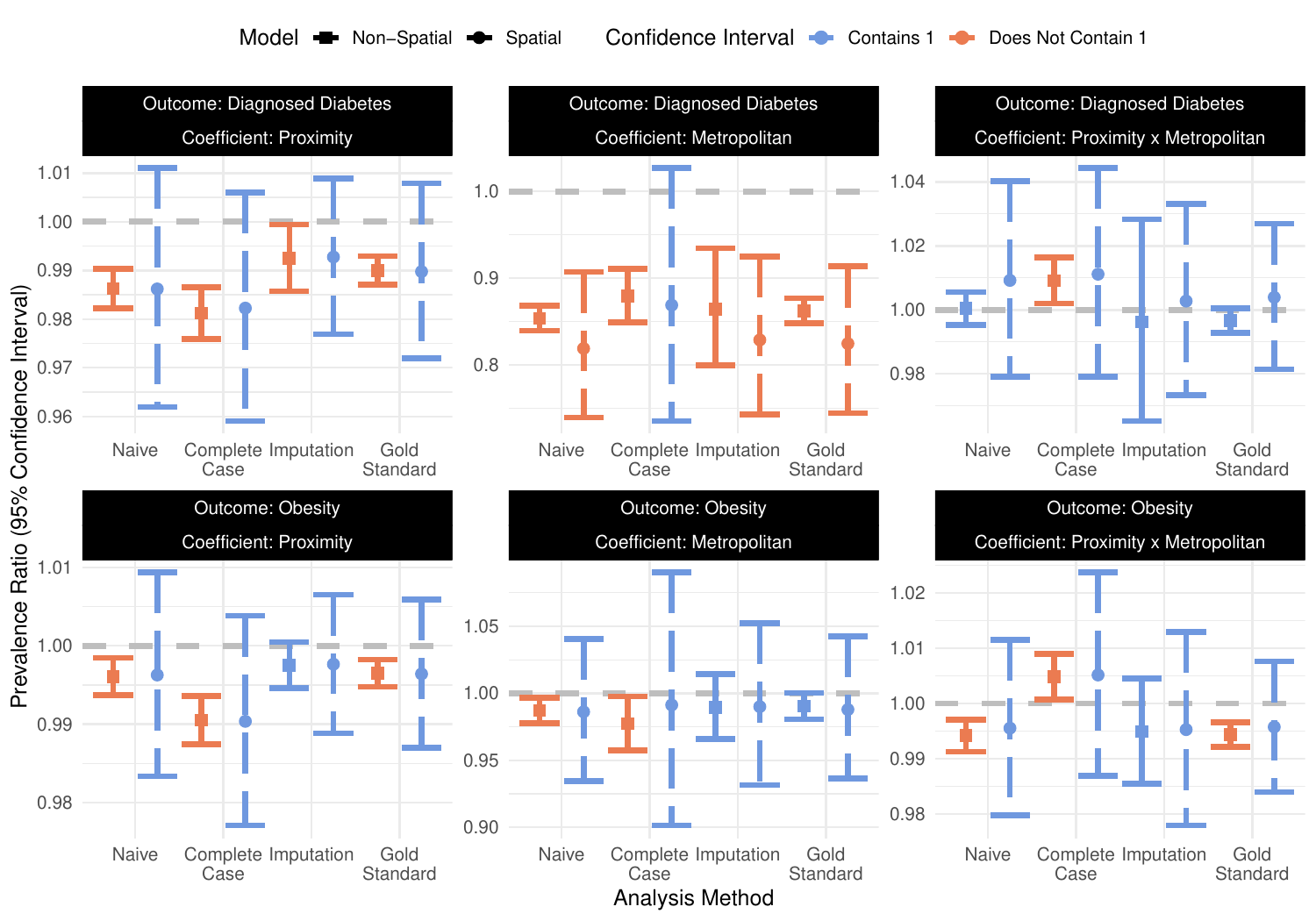}
\end{center}
\caption{Estimated adjusted prevalence ratios (with 95\% confidence intervals) for proximity to healthy foods, metropolitan status, and their interaction with the two health outcomes (diabetes and obesity) in the Piedmont Triad, North Carolina using four different analysis methods. Within each health outcome and method, estimates on the right with the dashed error bars came from the mixed-effects model allowing for spatial autocorrelation between bordering neighborhoods; estimates on the left with the solid error bars came from the non-spatial model assuming independence between neighborhoods.
\label{fig:forest_pr}}
\end{figure}

Notice that metropolitan status $Z$ and its interaction with proximity $X$, were included in the analysis model in Equation~\eqref{mod_prox_mixed}. This inclusion modified the interpretation of our parameters and broadened our interest from just the coefficient on $X$ to include the coefficient on the interaction $X \times Z$. Now, $PR_X= \exp(\beta_1)$ denotes the prevalence ratio between two \textit{non-metropolitan} neighborhoods whose proximities to healthy foods differed by $1$ mile, and $PR_{XZ} = \exp(\beta_3)$ denotes the ratio \textit{between} the prevalence ratios for a 1-mile proximity difference in metropolitan versus non-metropolitan census neighborhoods. While $PR_X$ and $PR_{XZ}$ were of primary interest, the prevalence ratio $PR_Z = \exp(\beta_2)$ between metropolitan and non-metropolitan tracts, adjusting for proximity, could also be helpful. Forest plots of the estimated adjusted prevalence ratios from each analysis method, along with their 95\% confidence intervals (95\% CI), are included in Figure~\ref{fig:forest_pr}. 

\subsubsection{Gold standard analysis}

Based on the gold standard analysis, worse access to healthy foods (i.e., larger distances $X$ to the nearest healthy food store) was associated with a slightly lower prevalence of diagnosed diabetes in non-metropolitan neighborhoods ($\widehat{PR}_X = 0.99$, 95\% CI: $0.97, 1.01$). That is, for every $1$ mile farther a neighborhood's nearest healthy foods retailer was from its population center, its prevalence of diagnosed diabetes was expected to decrease by $1\%$. However, the confidence interval captured the null ($PR_X = 1$). The same was found for metropolitan neighborhoods($\widehat{PR}_{XZ} = 1.00$, 95\% CI: $0.98, 1.03$). Rather than seeing worse access (i.e., farther proximity) associated with a higher prevalence of diabetes, as was previously seen, \cite{Ahern2011, Kanchi2021} our results were inconclusive about the direction of the effect. Importantly, our choice of proximity as an access measure, while those cited used density, captured a different facet of the food neighborhood environment. After adjusting for proximity, metropolitan neighborhoods were expected to have $18\%$ lower prevalence of diagnosed diabetes, on average ($\widehat{PR}_{Z} = 0.82$, 95\% CI: $0.74, 0.91$).

For obesity, the estimated prevalence ratio for proximity to healthy foods among non-metropolitan census tracts from the gold standard analysis was at the null ($\widehat{PR}_X = 1.00$, 95\% CI: $0.99, 1.01$). Estimates for metropolitan status ($\widehat{PR}_Z = 0.99$, 95\% CI: $0.94, 1.04$) and its interaction with proximity ($\widehat{PR}_{XZ} = 1.00$, 95\% CI: $0.98, 1.01$) were null, as well. The association between food access and obesity has seen conflict across the literature. Ahern et al. found that a higher density of some types of retailers (full-service restaurants and direct farm sales) led to lower rates, while other types (grocery stores) led to higher ones.\cite{Ahern2011} Meanwhile, others have found, like us, that there is not a relationship between them. \cite{White2007}

\subsubsection{Naive analysis}

Despite the errors in straight-line proximity, the naive estimates of prevalence ratios for proximity to healthy foods on diagnosed diabetes ($\widehat{PR}_{X^*} = 0.99$, 95\% CI: $0.96, 1.01$) and obesity ($\widehat{PR}_{X^*} = 1.00$, 95\% CI: $0.98, 1.01$) in non-metropolitan neighborhoods closely resembled the gold standard and led to the same conclusions about an absence of associations. (We use the subscript $X^*$ here to denote the naive estimates based on $X^*$ rather than $X$.) Again, the diagnosed diabetes prevalence was expected to be lower in metropolitan neighborhoods ($\widehat{PR}_{Z} = 0.82$, 95\% CI: $0.74, 0.91$), adjusting for proximity. In contrast, obesity prevalence was not expected to differ ($\widehat{PR}_{Z} = 0.99$, 95\% CI: $0.93, 1.04$). Estimates on the interaction between metropolitan status and proximity to healthy foods were near the null for both outcomes (diagnosed diabetes [$\widehat{PR}_{X^*Z} = 1.01$, 95\% CI: $0.98, 1.04$] and  obesity [$\widehat{PR}_{X^*Z} = 1.00$, 95\% CI: $0.98, 1.01$]).

\subsubsection{Complete case analysis}

As expected, deleting the $339$ unqueried census tracts led to the complete case analysis having wider confidence intervals for all coefficients and outcome models. In addition, despite its use of map-based proximity measures, the point estimates differed from the gold standard analysis. For instance, the estimated prevalence ratios for proximity among non-metropolitan census tracts on diagnosed diabetes ($\widehat{PR}_X = 0.98$, 95\% CI: $0.96, 1.01$) and obesity ($\widehat{PR}_X = 0.99$, 95\% CI: $0.98, 1.00$) were both farther from the null. Notably, the complete case analysis should be asymptotically consistent since $X$ is missing at random (MAR) for the unqueried neighborhoods, \cite{Little1992} but there are no such guarantees in a small sample like this one ($n=48$). These differences could also be explained by the higher estimated baseline prevalence, i.e., the predicted disease prevalence in a non-metropolitan neighborhood where proximity to healthy foods was $0$ miles (Supplemental Figure~S12). 

Again, the confidence intervals for the coefficient on the interaction captured the null for both outcomes (diagnosed diabetes [$\widehat{PR}_{XZ} = 1.01$, 95\% CI: $0.98, 1.04$] and obesity [$\widehat{PR}_{XZ} = 1.01$, 95\% CI: $0.99, 1.02$]), suggesting that the associations between proximity and disease prevalence were similar in metropolitan or non-metropolitan neighborhoods. Still, both estimates were farther from the null; for obesity, the estimate was even greater than one when the gold standard and naive estimates were less than or equal to it. Despite similar estimates on metropolitan status, the wider confidence interval led to a different conclusions about diagnosed diabetes ($\widehat{PR}_Z = 0.87$, 95\% CI: $0.73, 1.03$) in the complete case than the gold standard and naive analyses, but the same conclusion about obesity ($\widehat{PR}_Z = 0.99$, 95\% CI: $0.90, 1.09$).

\subsubsection{Imputation analysis}

In many ways, the imputation analysis fell between the naive, gold standard, and complete case analyses. The imputation estimates for both outcomes closely resembled those from the gold standard analysis and offered narrower confidence intervals than the complete case. For example, the estimated prevalence ratios for proximity to healthy foods in non-metropolitan neighborhoods were $\widehat{PR}_X = 0.99$ for diabetes (95\% CI: $0.98, 1.01$) and $\widehat{PR}_X = 1.00$ for obesity (95\% CI: $0.99, 1.01$). Still, we would not conclude that either health outcome was significantly associated with proximity to healthy foods among non-metropolitan neighborhoods.

The associations between proximity and the disease outcomes for metropolitan neighborhoods were similar to those for non-metropolitan ones, with $\widehat{PR}_{XZ} = 1.00$ for diagnosed diabetes (95\% CI: $0.97, 1.03$) and $\widehat{PR}_{XZ} = 1.00$ for obesity (95\% CI: $0.98, 1.01$). After adjusting for proximity, however, metropolitan neighborhoods were expected to have $17\%$ lower prevalence of diagnosed diabetes ($\widehat{PR}_Z = 0.83$, 95\% CI: $0.74, 0.92$) and similar prevalence of obesity ($\widehat{PR}_Z = 0.99$, 95\% CI: $0.93, 1.05$). Imputation's efficiency gains were seen for all coefficients, not just those relying on $X$. Therefore, unlike the complete case, the imputation analysis led to the same conclusions about all coefficients and outcomes as the gold standard.

\subsubsection{Spatial versus non-spatial models}

The most notable difference for the non-spatial models was smaller standard errors relative to the spatial models for either outcome and almost all analysis methods. Larger standard errors were expected for the mixed-effects models, as they capture the spatial autocorrelation between neighboring census tracts through the added random intercepts. With smaller standard errors in the non-spatial models, the efficiency gains for imputation over the complete case analysis depended on the outcome and coefficient, which was not seen for the spatial models where the gains were universal. 

Importantly, different conclusions would have been drawn from the non-spatial models (using nearly all analysis methods), as the smaller standard errors assuming independence between census tracts led to narrower 95\% CIs. For example, both health outcomes could be considered significantly associated with proximity to healthy foods based on the non-spatial gold standard analyses. Specifically, farther proximity (worse access) would be significantly associated with lower prevalences of diagnosed diabetes and obesity in non-metropolitan neighborhoods. For diagnosed diabetes, this same association would have been expected in metropolitan neighborhoods, while for obesity a stronger association would have been expected for metropolitan than non-metropolitan ones. These differences in conclusions demonstrate the importance of modeling spatial data accordingly, as ignoring the autocorrelation between neighboring census tracts could lead to incorrect conclusions about the relationships of interest.

\section{Conclusion}
\label{sec:conc}

Eating healthy food is a critical determinant of healthy living, yet many communities have inadequate access to fresh, nutritious food. Given the public health ramifications of disparities in food access and their potential connection to disparities in health, accurately identifying low-access, high-risk communities is a priority. Quantifying access requires an extensive number of distance calculations (e.g., between neighborhood population centers and healthy food stores), and these calculations are either (i) computationally simple methods but inaccurate or (ii) accurate methods but computationally complex. This trade-off is solved by adopting a partial validation study design and then devising a multiple imputation for measurement error approach to overcome the missing data. In doing so, our approach offers improved accuracy in quantifying food access and modeling its impact on health with computational ease. The proposed methods make large-scale analyses of accurate distance-based access measures feasible by overcoming the computational hurdles. Therefore, the geographic scope of access studies can expand to answer questions and drive decision-making for larger communities. 

Imputation was preferable over other model-based methods here for a few reasons. First, the covariate imputation framework can be used with various outcome models, whereas maximum likelihood estimators must be re-derived for any model changes. \cite{Tang2015} This flexibility was demonstrated in the real data analysis, where spatial correlation between disease prevalences in bordering neighborhoods was detected and incorporated into the analysis model with ease. Second, under correct analysis and imputation model specification, imputation offers consistent estimates in Poisson regression. Regression calibration is only approximately unbiased for nonlinear outcome models, although it has been found to work fairly well. \cite{Keogh2020}

In simulations and the application to the Piedmont Triad, imputation captured the benefits of the more accurate, map-based access measures despite these data being largely missing. Notably, the simulations  demonstrated that the proposed methods offered unbiased, often highly efficient estimates under either additive or multiplicative error mechanisms, making the methods broadly applicable to other modeling approaches for access and health. In the real data application, the imputation estimates were close to the gold standard and accompanied by narrower confidence intervals than the complete case. With the naive analysis shown to be biased toward or away from the null, the importance of accurately measuring access was also  illustrated. 

The Poisson regression model assumed equal dispersion (i.e., equal mean and variance) for the analysis model, which may not be reasonable in practice. Fortunately, it would be straightforward to adapt the multiple imputation procedure here to another model, like negative binomial regression, if over- or underdispersion were present. Another limitation stemmed from using the PLACES \cite{data_places} disease counts $Y$ that are SAEs (i.e., predictions from their own models) in the Piedmont Triad analysis. Though many other socioeconomic variables could be relevant to the analysis models here (e.g., poverty), they were not included as covariates since they were potentially already used to predict $Y$. Defining the origin for distance calculations from the census tracts posed one final challenge. A central point (the population center) was used to measure access for all residents, smoothing over the gradient of access within a census tract.

There are several interesting directions for future work. First, the subset of queried neighborhoods can be sampled based on any fully available information, providing an opportunity for strategic design. Perhaps existing optimal designs for Poisson regression \cite{Wang2006} or measurement error in other models \cite{amorim2021, lotspeichCJS} could be adapted. Second, map-based driving distance was used to gauge neighborhood proximity. With map-based driving time as another metric of food access that is expensive to obtain, it could be interesting to impute time-based food access from straight-line measures instead. Third, neighborhood rates of disease could also be related to \textit{unhealthy} food access, or whichever type of food (healthy or unhealthy) is more accessible. 




\bibliographystyle{wileyNJD-AMA}
\bibliography{Bibliography-MM-MC}





\section*{Acknowledgements}
The authors gratefully acknowledge the Andrew Sabin Family Center for Environment and Sustainability at Wake Forest University for a seed grant that supported this work. Computations were performed using the Wake Forest University (WFU) High Performance Computing Facility, a centrally managed computational resource available to WFU researchers including faculty, staff, students, and collaborators. 

\section*{Supplementary Materials}
\begin{itemize}
    \item \textbf{Additional appendices, tables, and figures:} The supplemental figures and tables referenced in Sections 2--5 are available online at \url{https://github.com/sarahlotspeich/food_access_imputation/blob/main/Supplementary_Materials.pdf} as Supplementary Materials.
    \item \textbf{R-package for imputation:} An \textsf{R} package \texttt{possum} that implements the multiple imputation methods described in this article is available at \url{https://github.com/sarahlotspeich/possum}.
    \item \textbf{R code and data for simulation studies and analysis:} The R scripts and data needed to replicate the simulation studies from Section~\ref{sec:sims} and  data analysis from Section~\ref{sec:app} are available at \url{https://github.com/sarahlotspeich/food_access_imputation}.
\end{itemize}


\end{document}